\documentclass[twocolumn]{article}
\usepackage{pkuarticle}
\AtBeginDocument{%
  }

\usepackage{array}
\usepackage{algorithm}
\usepackage{algpseudocode}
\usepackage{amsmath}
\usepackage{booktabs}
\usepackage{graphicx}  
\usepackage{multirow}
\usepackage{makecell}
\usepackage{subcaption} 
\usepackage{xspace}
\usepackage{xcolor}
\usepackage[most]{tcolorbox}
\usepackage{enumitem}
\usepackage{lipsum}
\usepackage[normalem]{ulem}
\usepackage{tabularx}
\usepackage{float}

\newcommand{\cmark}{\checkmark}

\definecolor{myback}{HTML}{efefef}
\definecolor{myframe}{HTML}{3e3a39}

\newtcolorbox[auto counter]{takeawaybox}{
  colback=myback,
  colframe=myframe,
  title=Takeaway~\thetcbcounter,
  fonttitle=\bfseries,
  boxrule=1pt,
  arc=1mm
}

\newcommand{\framework}{\texttt{DOPS}\xspace}
\newcommand{\platform}{NPU--PIM\xspace}

\newcommand{\rev}[1]{\textcolor{black}{#1}}

\newcommand{\highlightbibkeys}{}
\forcsvlist{\listadd\highlightbibkeys}{%
  wu2025pimoe,dettmers2022gpt3,xiao2023smoothquant,%
 huawei_mmad,huawei_nd2nz,huawei_aclformat,huawei_data_layout,huawei_transdata5hd%
}

\makeatletter
\let\old@lbibitem\@lbibitem
\def\@lbibitem[#1]#2{%
  \ifinlist{#2}{\highlightbibkeys}{\color{black}}{\color{black}}%
  \old@lbibitem[#1]{#2}%
}
\makeatother

\usepackage[numbers,sort&compress]{natbib}
\usepackage{etoolbox}

\renewcommand{\and}{,\ }
\title{Beyond Prefill-Decode Disaggregation: Dissecting LLM Inference for Heterogeneous Platforms via Dynamic Operator Scheduling}
\author{Jiaqi Yang \and Jiayi Li \and Yihan Fu \and Hongxiao Zhao \and Zhan Chen \and Qiuping Wu \and Yuchao Yang \and Bonan Yan}
\affiliation{Peking University, Beijing 100871, China\\
\small\texttt{jiaqiyang25@stu.pku.edu.cn,\ bonanyan@pku.edu.cn}}

\newcommand{\copyrightnotice}{%
  \begingroup
  \renewcommand{\thefootnote}{}%
  \footnotetext{\footnotesize
    This paper has been accepted for publication at the 59th IEEE/ACM
    International Symposium on Microarchitecture (MICRO). \copyright\ IEEE.
    Personal use of this material is permitted. Permission from IEEE must be
    obtained for all other uses, in any current or future media, including
    reprinting/republishing this material for advertising or promotional
    purposes, creating new collective works, for resale or redistribution to
    servers or lists, or reuse of any copyrighted component of this work in
    other works.}%
  \endgroup
}

\begin{document}
\maketitle
\copyrightnotice
\begin{abstract}
\textit{Prefill}--\textit{decode} disaggregation (\texttt{PD}) and roofline-based operator placement are common strategies for partitioning Large Language Model (LLM) inference across heterogeneous systems, but they are often insufficient in practice. End-to-end latency also depends on workload shape, runtime device contention, and persistent weight layout. We present \textbf{DOPS} (\uline{d}ynamic \uline{op}erator \uline{s}cheduling), a hardware-aware, closed-loop framework that jointly optimizes operator scheduling and blockwise weight layouts. DOPS constructs a stage-aware directed acyclic graph (DAG) and integrates two components: the \textsc{Bifocal} scheduler for dynamic operator-to-device placement and the Weight Layout Arbiter (WLA) for selecting hardware-efficient weight layouts under strict memory constraints. Across representative heterogeneous systems combining neural processing units (NPUs) and processing-in-memory (PIM) devices, \textsc{Bifocal} achieves geometric-mean speedups of 1.20$\times$ to 2.23$\times$ over the \texttt{PD} baseline. WLA provides an additional geometric-mean speedup of 1.28$\times$ to 1.33$\times$ over \texttt{Bifocal/Linear}. DOPS also supports systematic analysis of workload sensitivity and hardware scalability for LLM serving. The source code is available at \url{https://github.com/YIAI-02/TriForm}, and the visualization tool is demonstrated at \url{https://youtu.be/Ya_oMCyYno0}.
\end{abstract}

\section{Introduction}
\label{sec:intro}

Large Language Models (LLMs) have emerged as a cornerstone in artificial intelligence, rapidly evolving into versatile interfaces for conversational assistance, code generation, and information retrieval~\cite{openai2023gpt4,dakhel2023copilot,zhai2024llmir}. However, the serving cost and latency of LLM inference remain major barriers to broader deployment under tight service-level objectives (SLOs) and memory budgets~\cite{gong2025pastfuture,zhong2024distserve,oh2024exegpt,kim2025oaken,stojkovic2025dynamollm,wang2025step3}. This challenge is particularly acute on edge platforms constrained by limited power and area budgets~\cite{Park2024LPDDR_CXL_PNM,seo2025facil,xia2025kelle,yu2024cambriconllm,sun2025lincoln,liu2025ndpdimm,kim2025lia,Yun2024Duplex,na2025flexinfer}.

\begin{figure}[t]
    \centering
    \includegraphics[width=\linewidth]{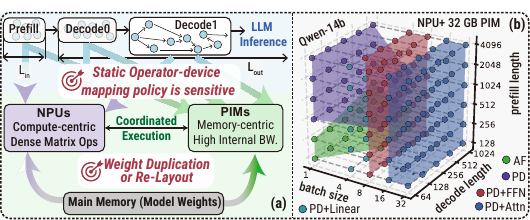}
    \caption{(a) Challenges for LLM inference on \platform systems. (b) The best static mapping policy for minimizing latency varies across workload configurations, parameterized by \{\textit{prefill} length ($x$), \textit{decode} length ($y$), and batch size ($z$)\}. 
\texttt{PD}, \texttt{AF}, \texttt{PD+Linear}, \texttt{PD+FFN}, and \texttt{PD+Attn} are different policies. See Table~\ref{tab:compared_methods} for details.}
    \label{fig:introduction}
\end{figure}

An LLM request consists of two stages with distinct execution profiles. \textit{Prefill} stage processes the full prompt with high arithmetic intensity, whereas \textit{decode} generates one token at a time while repeatedly streaming weights and the growing KV cache~\cite{kwon2023pagedattention,zhou2024survey,kim2025oaken}. Naturally, recent systems decouple these stages, since one compute-and-memory allocation rarely serves both efficiently~\cite{zhong2024distserve,patel2024splitwise,feng2025windserve}. This paradigm is known as \textit{prefill}--\textit{decode} disaggregation (\texttt{PD}).

This asymmetry makes heterogeneous architectures attractive. For edge deployment, NPU--PIM systems combine compute-centric neural processing units (NPUs), memory-centric processing-in-memory (PIM) devices, and persistent weight storage in main memory, as shown in Figure~\ref{fig:introduction}(a). NPUs excel at dense linear algebra through parallel multiply-accumulate arrays, but their efficiency depends on blocked execution aligned with hardware tile shapes~\cite{liao2021ascend,jouppi2023tpuv4}. On \textit{decode}-dominant workloads, even modern NPUs can become bandwidth-limited and underutilized~\cite{kwon2023pagedattention,heo2024neupims}. PIM devices complement NPUs by moving simple computation closer to memory and exploiting high internal bandwidth and bank-level parallelism for data-movement-heavy kernels~\cite{gu2025cent,Lee2022GDDR6_AIM,devic2022topimornot}. However, process and power constraints make PIM less suitable for compute-intensive kernels~\cite{quinn2025longsight,mahapatra2025instoragerag,liu2025heterrag,jang2025pimpal,han2025hybridbondedllm}.
\begin{figure*}[t]
    \centering
    \includegraphics[width=\linewidth]{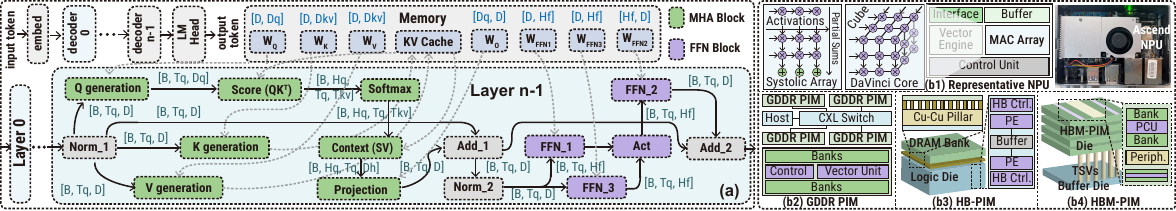}
    \caption{(a) Decoder-only transformer structure and operator-level dataflow within one layer. (b1) Representative NPU architecture. (b2)--(b4) Representative PIM architectures, including GDDR-based PIM (GDDR-PIM), hybrid-bonded PIM (HB-PIM), and HBM-based PIM (HBM-PIM).}
    \label{fig:preliminaries_transformer}
\end{figure*}

\textit{Prefill}--\textit{decode} disaggregation remains too coarse-grained for \platform execution. Prior work has explored finer-grained mappings, including attention-centric PIM offloading and static operator partitioning across NPUs and PIMs~\cite{park2024attacc,heo2024neupims,seo2024ianus}. 
{\color{black}To organize this design space, we classify operator-to-device schedulers along two axes: \textit{static} versus \textit{dynamic} placement, and \textit{online} versus \textit{offline} decision timing. Static placement uses fixed rules, whereas dynamic placement adapts to the workload and hardware. Online schedulers make decisions during serving, whereas offline schedulers make these decisions before deployment. Under diverse workloads and evolving \platform hardware, a fixed rule is unlikely to remain optimal.}
As illustrated in Figure~\ref{fig:introduction}(b), no single static policy dominates the workload space, even for a fixed model on a fixed hardware platform. The best operator-to-device mapping shifts with \textit{prefill} length, \textit{decode} length, and batch size. Policies that offload attention, linear, or FFN operators each dominate only part of the space, and the decision boundaries are irregular rather than monotonic. This variability makes universal mapping policies fundamentally unreliable.

Unlocking the full potential of heterogeneous computing platforms, therefore, requires answering fundamental questions: \textbf{Q1}: When is an \platform design genuinely beneficial for a target workload? \textbf{Q2}: How should operators be scheduled to achieve optimal performance under varying workload configurations?

Additionally, heterogeneous execution must consider how persistent weights are stored and accessed. In practice, the same weight block used by both NPU and PIM needs different physical data tiling organizations in memory. NPUs favor blocked layouts that efficiently feed tensor engines~\cite{liao2021ascend,jouppi2023tpuv4}, whereas PIM favors bank- or channel-aware layouts that align data placement with its internal parallelism~\cite{seo2025facil}. Maintaining a single device-agnostic copy forces repeated runtime relayout or repacking. Maintaining both device-specific copies in main memory eliminates that overhead, but quickly inflates the persistent memory footprint and can violate edge memory budgets~\cite{seo2024ianus,seo2025facil}. This raises another key question: \textbf{Q3}: How can persistent weight layouts be chosen to reduce relayout overhead without incurring the full memory cost of duplication?

To answer these questions, we present \framework, a hardware-aware framework for dissecting LLM inference for heterogeneous \platform systems. Our contributions include:

\begin{itemize}[leftmargin=*]
\item[$\diamond$] We formalize LLM inference on \platform as a coupled scheduling and data-storage optimization problem. We introduce \framework to jointly navigate this design space and maximize performance or efficiency.
\item[$\diamond$] We propose \textsc{Bifocal}, a dynamic scheduler that determines effective operator placement by combining rigorous finish-time estimation with downstream lookahead and reuse-aware biases. We further develop a Weight Layout Arbiter (WLA) that uses a two-stage block-coordinate search to select blockwise weight layouts in main memory.
\item[$\diamond$] We build a closed-loop workflow from user inputs through simulation, deployment, and validation. \framework achieves substantial speedups on representative \platform configurations. Beyond performance optimization, the workflow exposes system bottlenecks and supports hardware exploration across NPU-to-PIM ratios. We release the framework as open source to support further development and research.
\end{itemize}

We evaluate \framework on representative hardware configurations that combine a Huawei Ascend 910B NPU with SK Hynix AiM-based GDDR6-PIM devices. Our experiments show that the \textsc{Bifocal} scheduler achieves geometric mean speedups ranging from 1.20$\times$ to 2.23$\times$ over \texttt{PD} across different models. The proposed Weight Layout Arbiter (WLA) delivers an additional geometric mean speedup of 1.28$\times$ to 1.33$\times$ over \textsc{Bifocal}.

\section{Preliminaries}
\label{sec:Preliminaries}

\subsection{Transformer-based Large Language Models}
Figure~\ref{fig:preliminaries_transformer}(a) illustrates a representative decoder-only transformer architecture, which alternates multi-head attention (MHA) and feedforward network (FFN) blocks throughout the decoder stack.
Input tokens are mapped to embeddings and processed by $N$ decoder layers~\cite{vaswani2017attention}.
Each decoder layer contains multi-head self-attention and an FFN, together with normalization and residual connections. Grouped-query attention (GQA) and mixture-of-experts (MoE) designs further diversify the operator mix and memory traffic presented to the hardware~\cite{ainslie2023gqa,fedus2022switch,Yun2024Duplex}.
During both the \textit{prefill} and \textit{decode} stages, computation proceeds through all LLM layers.
The \textit{prefill} stage processes a prompt of length $L_{\mathrm{in}}$ and generates the initial KV cache.
The \textit{decode} stage generates $L_{\mathrm{out}}$ tokens autoregressively~\cite{kwon2023pagedattention,Sheng2023FlexGen}.
Here, $T_q$ denotes the number of query positions processed by the current attention operation, and $T_{kv}$ denotes the number of cached key-value positions visible to that operation.
Each \textit{decode} step usually has $T_q=1$, while attention reads an ever-growing cache of length $T_{kv}$.
The difference between the query and cache lengths produces distinct kernel regimes.
\textit{Prefill} is dominated by higher-intensity general matrix--matrix multiplication (GEMM).
\textit{Decode} is dominated by general matrix--vector multiplication (GEMV), KV-cache reads, and short kernels whose performance is sensitive to bandwidth and launch overhead~\cite{na2024llmcpu}.

Let $B$ denote the batch size, $D$ the model dimension, $D_q$ the query-projection width, $H_{\mathrm{KV}}$ the number of key/value heads, $D_h$ the head dimension, and $H_f$ the FFN hidden size.
During \textit{decode}, typical activation shapes include the query tensor $Q\!\in\![B,T_q,D_q]$, cached key and value tensors $K/V\!\in\![B,T_{kv},H_{\mathrm{KV}},D_h]$, and intermediate FFN activations of shape $[B,T_q,H_f]$, where $[\cdot]$ denotes a tensor shape.
\subsection{LLM Inference on \platform Hardware}

{\color{black}This paper focuses on \platform systems for edge LLM inference~\cite{xu2025llmnpu,lee2025aif}, where request shapes impose diverse compute and memory-access demands. NPUs and PIM devices provide complementary hardware substrates for these workloads.} As depicted in Figure~\ref{fig:preliminaries_transformer}(b1), a typical NPU combines a multiply-accumulate (MAC) array, vector units, and on-chip buffers for dense GEMM-style operators~\cite{Zhou2025Ascend,zhang2024llmcompass}. This organization is effective for dense GEMM kernels. However, operators with low arithmetic intensity can underutilize large matrix engines, making performance sensitive to on-chip buffer capacity and off-chip memory bandwidth~\cite{zhang2024llmcompass,heo2024neupims}.
As illustrated in Figure~\ref{fig:preliminaries_transformer}(b2)--(b4), PIM places lightweight compute logic close to or inside DRAM banks, reducing the cost of moving activations and KV-cache data across the memory interface~\cite{Zhao2024UMPIM,Park2024LPDDR_CXL_PNM}. Research prototypes and commercial systems have been demonstrated using GDDR-, HBM-, and LPDDR-based designs~\cite{Lee2022GDDR6_AIM,kim2021aquaboltxl,kwon2021fimdram,he2020newton,lee2021industrialpim,Park2024LPDDR_CXL_PNM}. More recent 2.5D and 3D integrations place compute logic under or near stacked memory and exploit through-silicon vias (TSVs), hybrid bonding, or monolithic 3D stacking to provide denser interconnects and higher effective bandwidth~\cite{Li2025H2LLM,Pan2025Stratum,bai2025accelstack,fu2025h2eal}. In the systems we target, NPUs and PIM devices access the same host-visible main-memory space. Persistent model weights are stored there and fetched through device-specific memory paths~\cite{seo2024ianus,Zhao2024UMPIM,seo2025facil}. This shared address space simplifies data sharing but does not eliminate layout mismatch because the devices prefer different physical tilings of the same weights.

\subsection{Weight Layout Mismatch for NPUs \& PIMs}
\texttt{Linear} denotes the default software-visible tensor organization, typically using row-major or column-major order, which preserves contiguous addressing and compatibility with host-side GEMM libraries~\cite{seo2025facil}. \texttt{NPU\_OPT} denotes an accelerator-friendly packed or fractal organization aligned with the target NPU's native compute granularity. On the Huawei Ascend NPU, low-level matrix multiplication uses operand-dependent fractal layouts such as ZZ, NZ, and ZN. \texttt{PIM\_OPT} denotes a bank-, channel-, and row-aware physical placement that aligns chunks or tiles with near-bank execution. This organization improves all-bank parallelism, bank-local computation, and row-buffer locality for GEMV~\cite{seo2024ianus}. Consequently, heterogeneous \platform systems may require relayout, duplication, or flexible address mapping for efficient data sharing across devices. Figure~\ref{fig:hw_and_layout}(a) illustrates how the three layouts map the same logical matrix to different memory locations.

\section{Motivation: Observations \& Problem Analysis}
\label{sec:motivation}
\textbf{Target problem}: solve the \textit{operator placement problem} for LLM inference on heterogeneous \platform systems, i.e., decide which device should execute each LLM operator to maximize performance or efficiency.
\begin{figure}[t]
	\centering
	\includegraphics[width=\linewidth]{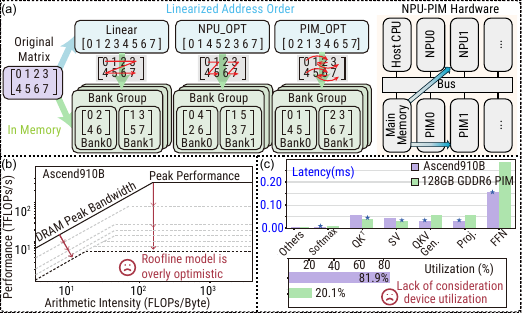}
	\caption{(a) Example of how Linear, NPU\_OPT, and PIM\_OPT layouts map the same logical matrix to bank groups of main memory. (b) The performance and bandwidth assumptions used in the roofline model are overly optimistic. (c) Roofline-model-based operator partitioning ignores device utilization.}
	\label{fig:hw_and_layout}
\end{figure}

\noindent
\textbf{Observations.}
\uline{The roofline model is insufficient for operator placement.}
Roofline-based heuristics overlook dynamic factors that are essential for accurate performance modeling. First, memory bandwidth is not a device-local constant. The effective bandwidth depends on inter-device transfers, heterogeneous memory paths and cache residency. Second, peak compute throughput often overestimates attainable performance, especially for nonlinear operators and short \textit{decode} kernels~\cite{chen2018tvm,zheng2020flextensor,zhang2024llmcompass,ghodrati2024tandem,na2024llmcpu}.  
Figure~\ref{fig:hw_and_layout}(b) shows that, on the Huawei Ascend 910B NPU, the roofline ``sweet spot'' shifts with workload size. When the workload is too small to saturate either the compute engine or the available memory bandwidth, the static roofline model becomes overly optimistic. These results indicate that classifying operators as compute-bound or memory-bound is incomplete; in practice, the effective roofline is inherently dynamic and workload-dependent.

\uline{A static placement rule lacks the time dimension of execution.} End-to-end computing latency depends on resource contention, queuing, overlap, and data residency, not only on the standalone latency of each operator~\cite{iliakopoulou2025chameleon,Jiang2025RAGO,stojkovic2025dynamollm,stojkovic2025tapas}. A policy that always chooses the locally faster device can send many ready operators to the same device, create a backlog on the critical path, and leave the other device underutilized, as shown in Figure~\ref{fig:hw_and_layout}(c) on the simulated heterogeneous Ascend910B-GDDR6 PIM platforms. In this setting, offloading operators to a device with slightly worse standalone latency can still shorten overall latency by reducing waiting time or avoiding a costly transfer on the queuing critical path. This issue is especially important in the LLM \textit{decode} stage, where token dependencies limit parallelism and make communication cost and device availability top priorities when assigning operators.

\uline{Weight tiling layout in main memory is coupled with operator placement.} Weight layout choice depends on whether a device can load data directly, whether relayout effort is required, and what effective load bandwidth is exposed to the runtime. An NPU\_OPT layout can improve tensor execution on the NPUs but increase the access cost for PIMs. A PIM\_OPT layout can improve bank-level parallelism in memory but introduce extra conversion before NPU execution~\cite{seo2024ianus,seo2025facil}. From a system perspective, once a weight block is reused across multiple layers or decode steps, its stored format changes both the effective load cost on each device and the likelihood that the block remains resident in a useful form. Therefore, the cost of a layout decision is history-dependent: it depends on which device consumed the block previously and which device is likely to consume it next. This temporal dependence is one of the main reasons why weight layout cannot be optimized independently from scheduling. Alternatively, duplicating multiple weight copies on each device (NPU and PIM) increases memory usage and violates edge memory budgets. 

\textbf{Motivation.} A quantitative framework is necessary to predict whether the heterogeneity of \platform systems is beneficial for a target LLM model, workload configuration, and hardware platforms. It should move beyond static roofline-model ceilings, search for a dynamic operator placement that is aware of communication and utilization, and jointly determine a weight layout that improves latency in both \textit{prefill} and \textit{decode} under memory constraints.

\section{\framework Framework}
\label{sec:framework}

\begin{figure}[t]
  \centering
  \includegraphics[width=\linewidth]{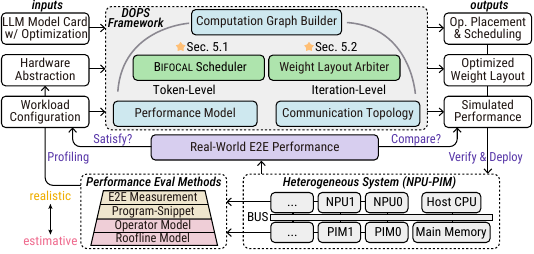}
  \caption{Overall workflow of \framework for dissecting LLM inference on a heterogeneous \platform system. E2E: end-to-end.}
  \label{fig:framework}
\end{figure}
To address the coupled scheduling and weight-layout problem identified in Section~\ref{sec:motivation}, we develop \framework, which organizes model, hardware, and workload information into a closed optimization loop. As shown in Figure~\ref{fig:framework}, \framework takes three classes of inputs: LLM model card with optimization (e.g. quantization, sparsification), hardware abstraction for the target \platform system, and workload configuration. From these inputs, \framework builds the computation graph, instantiates the performance model and communication topology, and produces three outputs: operator placement with the resulting execution schedule, optimized weight layouts in memory, and simulated end-to-end latency. The generated operator schedule and the optimized weight layouts are then deployed on the target \platform system, where measured execution is used to determine whether the target objectives are satisfied and to compare practical latency against simulated performance predictions. Detailed explanations are given as follows.

\subsection{Inputs}
\paragraph{LLM Model Card with Optimization.}
An LLM model card defines the target architecture and operator shapes, including variants that alter dependencies or tensor dimensions. \framework models quantization and sparsification because both affect bytes per operation, memory footprint, and communication traffic. Quantization specifies weight and activation precision, boundary casts, and per-channel or per-group scaling. Sparsification covers weight, activation, and attention sparsity, and activation sparsity may differ between \textit{prefill} and \textit{decode}. These abstractions cover representative inference schemes~\cite{Lee2024Tender,Ramachandran2025MicroScopiQ,Shin2025Grasp,zhao2024alisa}.

\paragraph{Hardware Abstraction.}
A hardware abstraction describes the target \platform system, including available device types, device counts, memory resources, and the interconnect.
\framework derives hardware-performance inputs from four abstraction levels: roofline models, operator-level measurements, program snippets, and end-to-end measurements (see the bottom-left ``performance eval methods'' block in Figure~\ref{fig:framework}). Among these, the roofline and operator models are directly consumed by \framework as scheduling inputs. 
To address the limitations discussed in Section~\ref{sec:motivation}, we apply device-specific corrections for utilization loss and short \textit{decode} kernels. The measurement-based operator model stores profiled per-operator latency for each device.
For multi-device configurations, the hardware model also specifies inter-device communication bandwidth and the network topology. These properties determine transfer paths, shared-link contention, and how much communication can overlap with computation.

\paragraph{Workload Configuration.}
A workload configuration describes how the target LLM is served and how broadly \framework explores the optimization space. On the serving side, it includes \textit{prefill} length $L_{\mathrm{in}}$, \textit{decode} length $L_{\mathrm{out}}$, batch size $B$, the optimization objective, and optional service-level constraints such as time to first token (TTFT), time per output token (TPOT), end-to-end latency, and memory capacity. On the optimization side, it specifies the search scope over layers, operators, devices, and formats, a candidate-evaluation budget, how far the scheduler looks ahead, and the termination criteria for exploration.

\subsection{Functional Components}

\paragraph{Computation Graph Builder.}
Using the inputs defined above, the Computation Graph Builder constructs a directed acyclic graph (DAG) that represents the target inference workload. The DAG is stage-aware and instantiates the operator sequence for both \textit{prefill} and \textit{decode} under the specified sequence lengths and batch sizes. Each DAG vertex represents a schedulable task.
{\color{black}The graph granularity is backend-defined. In the simplest case, one logical operator becomes one DAG vertex. When the backend exposes a fused kernel such as online softmax, kernel-level optimizations can be folded into one calibrated attention vertex. When sharding is enabled, a logical operator can be expanded into multiple shard vertices plus any required communication or reduction vertices.}
Each task vertex records the information needed by later stages, including tensor shape, precision, sparsity state, and arithmetic work. Directed edges encode both data dependencies and data movement, including activation transfers, KV-cache updates, weight loads from main memory, and any required relayout operation. This representation makes the coupling between operator placement and weight layout explicit within one DAG.

\paragraph{Performance Model.}
The performance model converts the annotated DAG into latency primitives used by the built-in optimization engines (Section~\ref{sec:2engines}). For each operator--device pair, it estimates compute time\footnotemark[1], memory-service time\footnotemark[1], and additional costs for weight loading, relayout, and cache-miss handling under the selected caching policy.
\footnotetext[1]{In this paper, we use the variable $\tau$ to represent time.}

\paragraph{Communication Topology.}
{\color{black}
The communication topology specifies the interconnect structure and bandwidth constraints. \framework currently supports two interconnect patterns: (i) a fully connected topology, where any pair of devices can communicate through a direct logical link; and (ii) a star topology, where device-to-device communication is routed through the host, while devices may also access host-side shared memory.
For either topology, \framework resolves each route and estimates the transfer latency of DAG data-movement edges using a link model consistent with AHEAD~\cite{abdelhafez2019ahead} and LogGP~\cite{alexandrov1995loggp}.

To model runtime contention, each resolved route is mapped to shared communication pipelines with availability times. A transfer starts only after its source data and required pipeline are ready. Concurrent transfers on the same pipeline are therefore serialized or delayed, whereas transfers on disjoint pipelines may overlap with one another and with computation. Tensor-parallel collectives, including scatter, reduce, and all-reduce, are inserted as communication vertices and evaluated with the same cost primitive.}
\begin{figure}[t!]
  \centering
  \includegraphics[width=\linewidth]{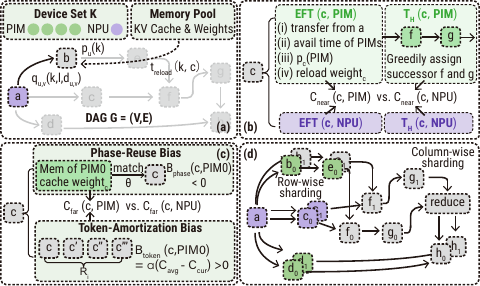}
  \caption{Overview of \textsc{Bifocal} scheduler. (a) A task DAG in \framework. (b) Example of Near-focus term $\mathrm{EFT}(v,k)$ and $\widehat{T}_H(v,k)$. (c) Example of Far-focus bias $B_{\mathrm{token}}(v,k)$ and $B_{\mathrm{phase}}(v,k)$. (d) Operator sharding and inserted communication tasks.}
  \label{fig:bifocal}
\end{figure}

\subsection{Built-in Optimization Engines}
\label{sec:2engines}
The optimization loop is driven by two engines: \textsc{Bifocal} scheduler and Weight Layout Arbiter (WLA).
More details are in Section~\ref{sec:Strategy}.

\textit{\textsc{Bifocal} scheduler} is a list scheduler that operates on the generated execution DAG to determine operator placement across the heterogeneous \platform system.
Inspired by the Heterogeneous Earliest Finish Time (HEFT) algorithm~\cite{Topcuoglu2002HEFT,arabnejad2013list,Khan2012CEFT,mcsweeney2020efficient}, we introduce an additional guidance metric, termed the ``bifocal score,'' to account for the coupled execution overhead between NPU and PIM.
The bifocal score comprises a near-focus score, which estimates the completion time of a ready operator given current device availability, and a far-focus score, which captures downstream effects such as device contention and cross-step reuse.

\textit{WLA} assigns storage formats to all weight blocks in main memory to minimize the predicted end-to-end latency. Each iteration evaluates a candidate layout by invoking \textsc{Bifocal} to estimate the makespan of the full inference DAG and to extract schedule-derived load statistics under that layout. It then alternates between one outer dominance-assignment update and several inner targeted-refinement updates to further reduce the predicted makespan.

\subsection{Outputs, Deployment, and Verification}
\label{subsec:outputs}

\paragraph{Operator Placement \& Scheduling.}
This specifies a concrete mapping from operators to devices (NPU0, NPU1, $\dots$, PIM0, PIM1, $\dots$) along with start time and completion time for every compute and communication event. It also records explicit communication events for dependency edges and models their potential pipelining with computation.

\paragraph{Optimized Weight Layout.}
The optimized weight layout is a blockwise storage plan for persistent weights in main memory. For each weight block, it records the selected layout format (Linear, NPU\_OPT, PIM\_OPT, etc.) and any remaining required conversion operations (e.g., \texttt{TransDataTo5HD} in the Huawei CANN framework~\cite{huawei_aclformat,huawei_data_layout,huawei_transdata5hd} or \texttt{tf.transpose()} in TensorFlow~\cite{tensorflow_transpose}) before execution on a given device.

\paragraph{Simulated Performance.}
\framework reports predicted performance for each candidate strategy under the same workload configuration and hardware settings. Because all candidates are evaluated under a common input specification, \framework can conveniently compare alternative scheduling policies, weight-layout strategies, and hardware configurations on an equal basis.

\paragraph{Verify $\&$ Deploy Loop.}
This loop moves the DAG schedules and the optimized weight layouts onto the target \platform system and measures real execution latency. Depending on the available hardware access and model fidelity, verification can be performed with program snippets or with full end-to-end latency. The measured results are used to check whether the target objectives are satisfied and to compare empirical latency against simulated predictions, which provides a direct assessment of model fidelity and optimization effectiveness.

\section{Optimization Algorithms}
\label{sec:Strategy}
This section explains the algorithms of the proposed two optimization engines. For each, we first formalize the underlying optimization problem, and then describe the methods adopted in \framework.

\subsection{\textsc{Bifocal} Scheduler} 
\label{subsec:bifocal}

\begin{algorithm}[tb]
\caption{\textsc{Bifocal} Optimizer\\(code is in \url{https://github.com/YIAI-02/TriForm})}
\label{alg:bifocal}
\small
\begin{algorithmic}[1]
\Statex \textbf{Input:} DAG $G=(V,E)$; legal-device sets $\{\mathcal{K}_v\}_{v\in V}$; weight format $\phi$; initial system state $s_0$.
\Statex \textbf{Output:} $\sigma:V\!\to\!\mathcal{K}$; $\tau=\{(\tau_{vs},\tau_{vc})\}_{v\in V}$; updated system state $s$.
\State Initialize $\sigma(v)\gets\bot$, $\tau_{vs}(v)\gets\bot$, and $\tau_{vc}(v)\gets\bot$ for all $v\in V$
\State Initialize $s\gets s_0$
\State Initialize ready set $\mathcal{R}\gets\{v\in V\mid \mathrm{Pred}(v)=\emptyset\}$
\While{$\mathcal{R}\neq\emptyset$}
  \State min-heap $\mathcal{Q}\gets\emptyset$
  \ForAll{$v\in\mathcal{R}$}
    \State $bestScore \gets +\infty$; $k_v \gets \bot$; $\widehat{\mathcal{H}}_v \gets \emptyset$
    \ForAll{$k\in\mathcal{K}_v$}
      \State $(score,\widehat{\mathcal H}_{v,k})\gets\textsc{Evaluate}(v,k,s,\phi)$
      \If{$score < bestScore$}
        \State $bestScore \gets score$; $k_v \gets k$; $\widehat{\mathcal{H}}_v \gets \widehat{\mathcal{H}}_{v,k}$
      \EndIf
    \EndFor
    \State Push $(bestScore,v,k_v,\widehat{\mathcal{H}}_v)$ into min-heap $\mathcal{Q}$
  \EndFor
  \State $(\sigma,\tau,s)\gets\textsc{Commit}(v^\star,k^\star,\widehat{\mathcal{H}}^\star,\phi,\sigma,\tau,s)$
  \State \textsc{Commit}$(v^\star,k^\star,\phi,\sigma,\tau,\mathrm{avail},c)$; \textsc{UpdateSystemState}
  \State Remove $v^\star$ from $\mathcal{R}$ and insert any newly ready tasks into $\mathcal{R}$
\EndWhile
\end{algorithmic}
\parbox[t]{0.98\linewidth}{$\phi$ denotes the weight-format map. \textsc{Evaluate} returns the score in Equation~\ref{eq:bifocal_score} and its associated tentative hint map. The superscript ${}^\star$ marks the minimum-key candidate popped from $\mathcal{Q}$.}
\end{algorithm}

\subsubsection{Problem Formulation}
As shown in Figure~\ref{fig:bifocal}(a), let $G=(V,E)$ be the execution DAG instantiated by the Computation Graph Builder, with a unique source (starting) vertex and a unique sink (end) vertex.
A logical operator may be decomposed into multiple shard vertices $v\in V$, which may be placed on different devices.
Each edge $(u,v)\in E$ (from vertex $u$ to $v$) enforces precedence and carries a data payload $d_{uv}>0$.
Let $\mathcal{K}=\{1,\dots,K\}$ be the device set (e.g., \{NPU0, NPU1, $\dots$, PIM0, PIM1, $\dots$\}), and let $\mathcal{K}_v\subseteq\mathcal{K}$ denote the legal devices for vertex $v$.
For a vertex $v$ and legal device $k\in\mathcal{K}_v$, the performance model returns an estimated device-dependent execution latency $p_v(k)$.
For an edge $(u,v)\in E$ and device pair $(k,\ell)$, the performance model returns an estimated data-transfer latency $q_{uv}(k,\ell,d_{uv})$.
For a weight-bearing operator, let $weight_v$ denote its dominant reusable weight block. For a non-weight-bearing vertex, let $weight_v=\varnothing$ and define $t_{\mathrm{reload}}(\varnothing,k,c)=0$.

{\color{black}
During optimization, \textsc{Bifocal} maintains system state $s$, which includes ready tasks, device availability, cache state $c$, near-future hints $\mathcal{H}$, and cross-step weight-device hints $\theta$. $\mathcal{H}:V\rightharpoonup\mathcal{K}$ stores tentative placements within the current lookahead window; after each commit, stale entries are dropped and remaining hints are refreshed. A schedule maps $\sigma:V\rightarrow\mathcal{K}$ with $\sigma(v)\in\mathcal{K}_v$ and start/completion times $(\tau_{vs},\tau_{vc})$.
}

A feasible schedule satisfies the following precedence and resource constraints:
\begin{equation}
\begin{aligned}
  \tau_{vs} &\ge \tau_{uc}+q_{uv}\bigl(\sigma(u),\sigma(v),d_{uv}\bigr),
  && \forall (u,v)\in E,\\
  \tau_{vs} &\ge a_{\sigma(v)}(v),
  && \forall v\in V,\\
  \tau_{vc} &= \tau_{vs}+p_v\bigl(\sigma(v)\bigr)
  +t_{\mathrm{reload}}\bigl(weight_v,\sigma(v),c_v\bigr),
  && \forall v\in V.
\end{aligned}
\label{eq:constraints}
\end{equation}
where $a_k(v)$ is the value of $\mathrm{avail}(k)$ when $v$ is evaluated, and $c_v$ is the cache state immediately before $v$ starts. The first line enforces data dependencies, the second enforces device availability, and the third defines completion time as the start time plus device execution time and any cache-induced weight-reload penalty.

The overall objective is to minimize the makespan $T=\tau_{tc}$ ($t$ is the sink vertex) by finding an operator-to-device mapping $\sigma$.

\subsubsection{\textsc{Bifocal} Optimizer}

\textsc{Bifocal} is a list scheduler that incrementally constructs an operator placement and execution timeline for an LLM inference DAG.
Representative scheduling steps are illustrated in Figure~\ref{fig:bifocal}(b--c).
In the simplified example in Figure~\ref{fig:bifocal}(b--c), each LLM operator (e.g., $\mathbf{QK}^\top$) corresponds to one task and is the minimum allocation unit.
Algorithm~\ref{alg:bifocal} summarizes the main control flow. We call a vertex a ``\emph{ready task}'' if all its predecessor vertices have been committed to the current partial schedule.
At each iteration, the scheduler re-evaluates every ready task under the current partial schedule, identifies its best legal device, and pushes the resulting candidate into a min-heap keyed by the bifocal score.
It then commits the ready task whose best legal placement has the minimum score, updates the partial timeline and cache state, and refreshes both hint maps. 

{\color{black}We define
\begin{equation}
\begin{aligned}
\mathrm{Score}(v,k) &= C_{\mathrm{near}}(v,k)+C_{\mathrm{far}}(v,k), \\
C_{\mathrm{near}}(v,k) &= (1-\gamma)\,\mathrm{EFT}(v,k)+\gamma\,\widehat{T}_H(v,k),\\C_{\mathrm{far}}(v,k) &= B_{\mathrm{phase}}(v,k)+B_{\mathrm{token}}(v,k).
\end{aligned}
\label{eq:bifocal_score}
\end{equation}
where $\gamma\in[0,1]$ is a fixed scheduler hyperparameter and smaller scores are preferred.}
The individual terms are defined as follows:

\textbf{Near-focus earliest finish time.} The term $\mathrm{EFT}(v,k)$ is the local completion-time estimate if $v$ is placed on device $k$ under the current partial schedule, and $\mathrm{EST}(v,k)$ is the corresponding earliest start time. Let $\mathrm{Pred}(v)$ denote the set of predecessor vertices of $v$.
It is computed from (i) the maximum predecessor-arrival time, where each arrival combines predecessor completion and the corresponding transfer latency $q_{uv}(\cdot)$, (ii) the earliest availability time of device $k$, (iii) the device-dependent execution time $p_v(k)$, and (iv) the weight-reload penalty induced by the current cache state $c$. Concretely, we use
\begin{equation}
 \begin{aligned}
  \mathrm{EST}(v,k)
  &= \max\!\Bigl(\mathrm{avail}(k),\ \max_{u\in \mathrm{Pred}(v)}\bigl[\tau_{uc}+q_{uv}\bigl(\sigma(u),k,d_{uv}\bigr)\bigr]\Bigr),\\
  \mathrm{EFT}(v,k)
  &= \mathrm{EST}(v,k)+p_v(k)+t_{\mathrm{reload}}\bigl(weight_v,k,c\bigr).
\end{aligned}   
\end{equation}

For a source vertex, we use the convention $\max\emptyset=-\infty$, so its earliest start time is $\mathrm{avail}(k)$. All terms have units of time, and lower values are preferred.

\textbf{Near-focus DAG-window term} ($\widehat{T}_H(v,k)$) approximates the downstream impact of committing $(v,k)$ by exploring a short successor chain containing at most $H$ vertices.
The window is formed by recursively following the outgoing dependency with the largest payload until it reaches $H$ vertices or no successor remains.
The \textsc{Bifocal} scheduler performs a lightweight lookahead simulation over this window. It fixes the root placement of $v$ on $k$, uses entries from $\mathcal{H}$ as consistency hints for overlapping vertices, enumerates legal NPU/PIM device-type assignments for the remaining window vertices, and selects the assignment with the smallest simulated completion time.
{\color{black}
The routine returns $\widehat{T}_H(v,k)$ and a tentative window mapping; after the selected task is committed, this mapping is saved in $\mathcal{H}$ only as reusable near-future hints.}

\textbf{Far-focus phase-reuse bias} ($B_{\mathrm{phase}}(v,k)$) captures schedule-local weight--device consistency and \textit{prefill} residency potential.
In addition to near-future vertex placements, the \textsc{Bifocal} scheduler maintains the weight--device hint map $\theta$ defined above.
A candidate whose device type differs from $\theta(weight_v)$ incurs a positive penalty proportional to the estimated reload overhead; a matching candidate incurs no such penalty.
During \textit{prefill}, we additionally give a negative bias to a device class whose aggregate capacity can retain the model's full weight set, scaled by the estimated reload time.

\textbf{Far-focus token-amortization bias} ($B_{\mathrm{token}}(v,k)$) captures the autoregressive nature of decoding.
During \textit{decode}, let $R_i=L_{\mathrm{out}}-i+1$ denote the effective remaining \textit{decode} horizon at step $i$.
We compare the current-token cost of choosing device $k$ with the estimated average per-token cost if the same placement is reused for the remaining \textit{decode} tokens:
\begin{equation}
B_{\mathrm{token}}(v,k)
  = \alpha\bigl(
  \widehat{C}_{\mathrm{avg}}(v,k;i,L_{\mathrm{out}},c)
  - \widehat{C}_{\mathrm{cur}}(v,k,c)
  \bigr),
\end{equation}
where $\alpha\ge 0$ is the token-amortization scaling coefficient, $\widehat{C}_{\mathrm{cur}}$ is the current-token cost under the present cache state, and $\widehat{C}_{\mathrm{avg}}$ is the corresponding average per-token cost after spreading any one-time migration or reload overhead across the remaining horizon.
Thus, when many \textit{decode} tokens remain, the \textsc{Bifocal} scheduler may accept a higher current-token cost if the placement is expected to reduce the average cost of later steps.

Once the best assignment has been determined, the scheduler commits the vertex and updates the scheduling state, device availability, the cache state $c$, the near-future hint map $\mathcal{H}$, and the schedule-local weight--device hint map $\theta$.

\subsubsection{A Representative Example}
Figure~\ref{fig:bifocal}(b--c) illustrates how \textsc{Bifocal} selects a placement for one vertex.
For a vertex $c$, which reuses $weight_c$, we compare running it on NPU versus PIM0.
We first compute the near-focus scores $C_{\mathrm{near}}(c,\mathrm{PIM0})$ and $C_{\mathrm{near}}(c,\mathrm{NPU})$ to determine the preferable short-term placement under the current partial schedule (Figure~\ref{fig:bifocal}(b)).
This comparison combines the immediate earliest-finish estimate $\mathrm{EFT}(c,k)$ with the short DAG-window lookahead $\widehat{T}_H(c,k)$, so it reflects not only when $c$ itself would finish, but also how committing $c$ to device $k$ perturbs a short successor chain.
We then evaluate the far-focus score (Figure~\ref{fig:bifocal}(c)).
For example, matching $\theta(weight_c)$ avoids the device-change penalty in $B_{\mathrm{phase}}(c,\mathrm{PIM0})$; during \textit{prefill}, this term can also be negative when the aggregate PIM capacity can retain the model's full weight set.
Then, during a \textit{decode} step with a long remaining horizon $R_i$, the scheduler can deliberately choose PIM0 and accept a higher current migration cost to achieve a lower average cost $\widehat{C}_{\mathrm{avg}}$ for future \textit{decode} tokens. The token-amortization bias $B_{\mathrm{token}}$ captures this effect.
These far-focus terms can outweigh another device's near-focus advantage, causing the scheduler to prefer a placement with lower long-term cost even when its immediate $C_{\mathrm{near}}$ is higher.

Figure~\ref{fig:bifocal}(d) illustrates two forms of operator sharding.
Row-wise sharding creates mutually independent shards, so the scheduler may interleave shard execution with other ready tasks and establish new DAG edges directly from each shard.
Column-wise sharding, in contrast, requires explicit synchronization and reduction; therefore, communication tasks such as \texttt{reduce}, \texttt{all-reduce}, \texttt{gather}, or \texttt{scatter} are inserted into the DAG and scheduled explicitly.
{\color{black}
The graph-construction interface is extensible: additional serving-parallelism modes can be represented by inserting the corresponding vertices and dependency constraints.}
For MoE models, we insert router vertices to model token-to-expert routing explicitly, including the selection of the top-$k$ experts (FFN blocks) for each token. The resulting dispatch and combine dependencies are materialized as ordinary DAG edges or communication vertices.
Task partitioning, KV-write pinning, router insertion, and communication-vertex insertion are all completed during DAG preprocessing.

\begin{table*}[tb]
	\centering
	\small
\caption{Operator-placement policies evaluated on heterogeneous computing platforms}
	\begin{tabularx}{\textwidth}{>{\centering\arraybackslash}m{1.5cm} | >{\centering\arraybackslash}m{4.2cm} | >{\raggedright\arraybackslash}m{2.5cm} |  >{\raggedright\arraybackslash}X}
		\noalign{\hrule height 1pt}
		\textbf{Method} & \textbf{Category} & \hfil \textbf{\textit{prefill} placement} & \hfil \textbf{\textit{decode} placement} \\
		\hline
			\texttt{PD}& Static, \textit{prefill}--\textit{decode} disaggregation &
		All operators on NPU. &
		All operators on PIM. \\
		\hline
		\texttt{AF} & Static, attention-FFN disaggregation &
		\multicolumn{2}{m{10cm}}{Attention-related operators on PIM in both stages. Remaining operators on NPU.} \\
		\hline
		\texttt{PD+Linear} & FACIL-inspired static rule~\cite{seo2025facil}&
		All operators on NPU. &
		Linear operators compiled to GEMV on PIM. Other operators on NPU. \\
		\hline
		\texttt{PD+Attn} & AttAcc-inspired static rule~\cite{park2024attacc} &
		All operators on NPU. &
		Attention-related operators on PIM. Other operators on NPU. \\
		\hline
		\texttt{PD+FFN} & IANUS-inspired static rule~\cite{seo2024ianus} &
		All operators on NPU. &
			Fully connected (FC) operators on PIM during \textit{decode}. Remaining operators on NPU. \\
		\hline
			\textbf{\textsc{Bifocal}} & \textbf{Dynamic} (\textbf{this work}) &
			\multicolumn{2}{m{10cm}}{\textbf{\textit{Prefill}--\textit{decode} joint optimization.} A dynamic \textsc{Bifocal} scheduler with a bifocal score.} \\
		\noalign{\hrule height 1pt}
	\end{tabularx}
	\label{tab:compared_methods}
\end{table*}

\begin{table}[tb]
	\centering
	\small
		\caption{Hardware settings and performance models}
	\begin{tabularx}{\columnwidth}{>{\centering\arraybackslash}m{2.4cm} | >{\centering\arraybackslash}m{2.4cm} | >{\centering\arraybackslash}X}
		\noalign{\hrule height 1pt}
		\textbf{Type} & \textbf{NPU}  &  \textbf{PIM}  \\
		\hline
		\textbf{Device} &
		Huawei Ascend 910B&
		SK Hynix AiM device\\
		\hline
			\textbf{Peak compute throughput} &
			280 TFLOPS @ FP16 &
			16 TFLOPS @ FP16\\
		\hline
		\textbf{Memory} &
		16 GB &
			16 GB/device \\
		\hline
			\textbf{Peak memory bw.} &
		0.8 TB/s &
			\hfil 16 TB/s/device \\
		\hline
			\textbf{Scheduling input} &
		{\color{black}Calibrated roofline model} &
		Operator model \\
		\hline
			\textbf{Verification model} &
		Program snippets &
			{\color{black}Program snippets with the AiM simulator~\cite{gu2025cent}} \\
		\noalign{\hrule height 1pt}
	\end{tabularx}
	\label{tab:hardware}
\end{table}

\subsection{Weight Layout Arbiter}
\label{subsec:weight_Layout_Arbiter}

\begin{figure}[tb]
	\centering
	\includegraphics[width=\linewidth]{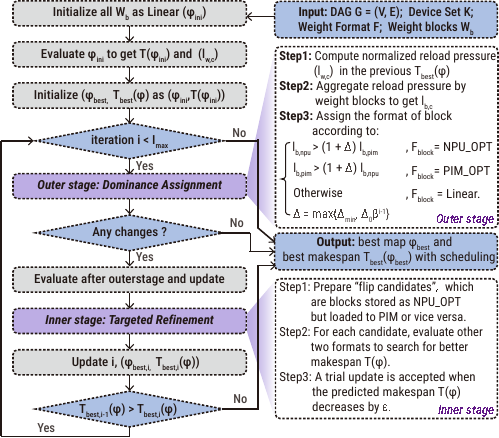}
		\caption{Two-stage block-coordinate search for optimizing the blockwise weight-format map. The left panel summarizes initialization; the upper-right and lower-right panels specify the outer- and inner-stage update rules, respectively.}
	\label{fig:two_stage_bcd}
\end{figure}

\subsubsection{Problem Formulation}

The WLA optimizes the blockwise assignment of weight-storage formats for a fixed workload and hardware configuration.
It shares the same DAG $G=(V,E)$, device set $\mathcal{K}$, and cost primitives as the \textsc{Bifocal} scheduler.

Let $W$ be the set of weight identifiers referenced by $G$.
Weights are partitioned into stable blocks $B$, where each block $b\in B$ corresponds to a subset $W_b\subseteq W$.
Let $F$ be the supported storage format set, including $\mathrm{Linear}$, $\mathrm{NPU\_OPT}$, and $\mathrm{PIM\_OPT}$ as described in Section~\ref{sec:Preliminaries}.
A blockwise assignment is a map $\phi:B\rightarrow F$, inducing a per-weight map $\varphi(w)=\phi(b)$ for all $w\in W_b$.
For any map $\phi$, let $T(\phi)$ denote the makespan predicted by \textsc{Bifocal} for the schedule induced by $\phi$. The optimization problem is $\min_{\phi:B\rightarrow F} T(\phi)$.

\subsubsection{Method: Two-Stage Block Coordinate Search Strategy}
\label{sec:bg_layout_mismatch}

We solve the above problem using a two-stage block-coordinate search strategy that is tailored to the implementation cost of repeated schedule evaluation.
The search still operates over the discrete Cartesian product $F^{|B|}$, but instead of performing a full exact coordinate sweep, it alternates between an outer dominance assignment stage and an inner targeted refinement stage.

Figure~\ref{fig:two_stage_bcd} summarizes the overall loop on the left and the two update stages on the right. We initialize $\phi_{\mathrm{ini}}$ by assigning $\mathrm{Linear}$ to all blocks, and evaluate the schedule once using \textsc{Bifocal} to obtain $T(\phi_{\mathrm{ini}})$ together with schedule-derived per-weight load statistics. Specifically, for each weight identifier $w\in W$ and device class $c$, let $\ell_{w,c}$ denote the load pressure induced by the most recently evaluated schedule of $w$ on device class $c$. WLA aggregates these per-weight statistics to block granularity.
The resulting block-level quantities are then normalized to obtain $\tilde{\ell}_{b,\mathrm{NPU}}$ and $\tilde{\ell}_{b,\mathrm{PIM}}$, which guide the dominance decision in the outer stage.

\paragraph{Outer Stage: Dominance Assignment.}
This stage performs a coarse blockwise reassignment based on the block-level reload pressure observed in the previous schedule. For each block $b$, the dominance relation between $\tilde{\ell}_{b,\mathrm{NPU}}$ and $\tilde{\ell}_{b,\mathrm{PIM}}$ is then used to determine a promising format for $b$, as illustrated in Figure~\ref{fig:two_stage_bcd}.

\paragraph{Inner Stage: Targeted Refinement.}
This stage refines the coarse map through focused local coordinate search. It first builds a candidate set containing the blocks whose current format still disagrees with the observed loading behavior in the refreshed schedule. For each candidate block $b$, the method holds all other block assignments fixed, explicitly evaluates the neighboring maps $\phi^{(b\leftarrow f)}$ for $f\in F\setminus\{\phi(b)\}$, and accepts a flip only when the predicted makespan decreases by more than $\epsilon$, as illustrated in the lower-right panel of Figure~\ref{fig:two_stage_bcd}. Each accepted flip therefore improves the current solution within the one-block neighborhood explored at that step. After each accepted flip, the arbiter invokes \textsc{Bifocal} again before processing the next candidate.

The procedure ends when the outer stage does not produce further format changes, when the iteration budget is reached, or when the tolerance criterion of early-stop $\epsilon$ is activated. The final output is the best blockwise map $\phi_{\mathrm{best}}$ and its predicted makespan $T_{\mathrm{best}}$.

\section{Experiments}
\label{sec:Experiments}


\begin{figure*}[tb]
  \centering
  \includegraphics[width=\linewidth]{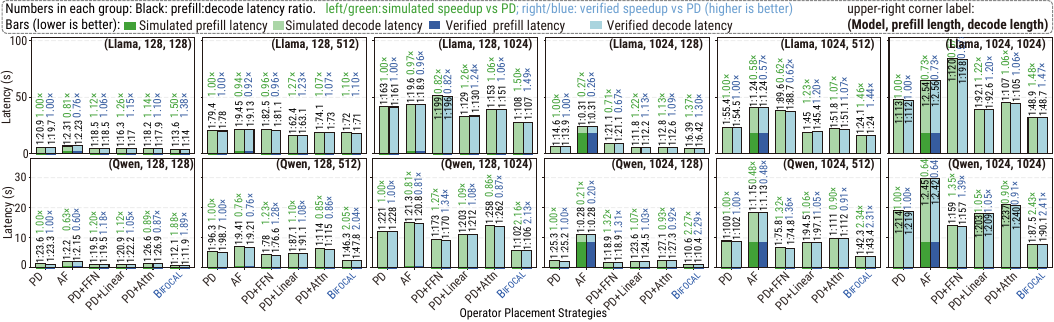}
  \caption{\color{black}Verification of \framework on \texttt{HP32} for representative workloads.}
  \label{fig:exp_verification_npu}
\end{figure*}

\begin{figure}[tb]
  \centering
  \includegraphics[width=\linewidth]{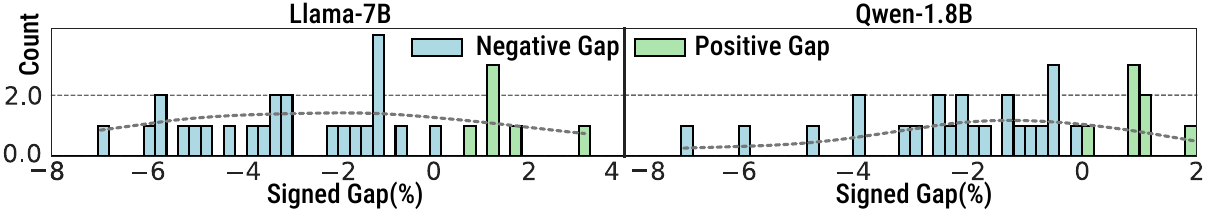}
  \caption{\rev{Distribution of the signed percentage gap between simulated and verified end-to-end speedup for Llama-7B and Qwen-1.8B.}}
  \label{fig:exp1_histogram}
\end{figure}

\subsection{Setup}

\textbf{Benchmarks and workloads.} We evaluate \framework on a representative set of models, including Llama-7B, Llama-13B, Llama-70B, Qwen-1.8B, Qwen-7B, Qwen-14B, and Mixtral-8$\times$7B, covering dense architectures, GQA, and MoE models. {\color{black}For Mixtral-8$\times$7B, the graph builder instantiates 8 local experts with deterministic top-2 routing.} For the scheduling experiments in Sections~\ref{subsec:exp1} and~\ref{subsec:exp2}, we sweep \textit{prefill} lengths over \{128, 512, 1024, 2048\}, \textit{decode} lengths over \{128, 256, 512, 1024\}, and batch sizes over \{1, 4, 8, 16\}. For the weight-layout study in Section~\ref{subsec:exp3}, we sweep \textit{prefill} and \textit{decode} lengths over \{8, 32, 64, 128, 1024\}, while emphasizing the shorter settings because relayout overhead is amortized more effectively over long sequences~\cite{seo2025facil}. 
{\color{black}To prevent the evaluation from being dominated by repeated weight swapping over the NoC, each target hardware configuration is chosen so that it can accommodate the full model weights under the evaluated workload settings. Accordingly, Llama-70B and Mixtral-8$\times$7B use INT8 quantization~\cite{dettmers2022gpt3,xiao2023smoothquant}, while the remaining models use FP16.}
We report end-to-end latency together with its \textit{prefill} and \textit{decode} components. Due to space constraints, we present only representative cases in the paper; the full sweep is available in our \url{https://github.com/YIAI-02/TriForm}.

\textbf{Hardware settings.} We use a Huawei Ascend 910B NPU and SK Hynix AiM devices based on GDDR6-PIM~\cite{gu2025cent}. We construct four hybrid-platform variants by pairing the NPU with 0, 2, 4, and 8 AiM devices, denoted as \texttt{HP0}, \texttt{HP32}, \texttt{HP64}, and \texttt{HP128}, respectively, where the suffix denotes the total PIM capacity in GB. The interconnect is a fully connected PCIe Gen4 x16 fabric. Both the NPU and the PIM devices fetch weight blocks from main memory. When the stored layout of a weight block does not match the target device, loading is followed by a format conversion. We include this overhead in the reported end-to-end latency. By default, the KV cache is stored in PIM at head granularity. Loaded weight blocks remain resident while capacity permits; otherwise, they are evicted according to an LRU policy. We assume maximal overlap between computation and weight loading. Table~\ref{tab:hardware} summarizes the hardware parameters together with the models used for scheduling and verification.

{\color{black}
For the NPU scheduling input, we use a calibrated roofline model. For each NPU sub-kernel $s$ with FLOPs $f_s$ and memory traffic $m_s$, DOPS estimates its latency as
\[
\tau_s =
\max\!\left(
\frac{f_s}{P_{e(s)}\cdot \max(u(f_s),\epsilon_u)},
\frac{m_s}{B_{e(s)}}
\right) +\tau_{\mathrm{kernel},e(s)},
\]
where $e(s)\in \{{\mathrm{Cube},\mathrm{Vector}}\}$ denotes the Ascend execution engine, $P_{e(s)}$ and $B_{e(s)}$ are engine throughput and memory bandwidth, $u(f_s)$ is a fitted saturated logistic function, $\epsilon_u$ stabilizes utilization, and $\tau_{\mathrm{kernel},e(s)}$ is CANN-profiled launch overhead.

}

In addition, \framework supports LLMCompass~\cite{zhang2024llmcompass} and our in-house Ascend 910B simulator; the latter supports configurable operator splitting and hybrid execution. For weight-relayout modeling, we follow the methodology in~\cite{seo2025facil} and estimate the memory-side relayout cost using Ramulator2~\cite{luo2023ramulator2}; NPU-side format-conversion latency is estimated using Huawei CANN conversion kernels~\cite{huawei_nd2nz}.

\textbf{Baselines.} For Section~\ref{subsec:exp1}, we compare \textsc{Bifocal} with the static policies summarized in Table~\ref{tab:compared_methods}: the \texttt{PD} and \texttt{AF} baselines and the prior-work-inspired \texttt{PD+Attn}, \texttt{PD+FFN}, and \texttt{PD+Linear} rules. For a fair comparison, we reimplement only each method's operator-to-device placement rule on a common \platform target. We do not import hardware-specific datapaths, memory-system modifications, or runtime mechanisms from the original systems.

For Section~\ref{subsec:exp3}, we compare five layout strategies: \texttt{PD/Linear}, \texttt{PD/Dual}, \texttt{Bifocal/Linear}, \texttt{Bifocal/WLA}, and \texttt{Bifocal/Dual}. The \texttt{Dual} variants maintain separate \texttt{NPU\_OPT} and \texttt{PIM\_OPT} copies. Neither device incurs runtime format-conversion overhead, but persistent storage doubles. {\color{black}We use \texttt{Bifocal/Dual} as a reference upper bound under idealized dual-format access, where each device reads weights in its preferred physical layout.} \texttt{Bifocal/WLA} combines the \textsc{Bifocal} scheduler in Section~\ref{subsec:bifocal} with the WLA mechanism in Section~\ref{subsec:weight_Layout_Arbiter}.

{\color{black}
For Sections~\ref{subsec:exp1}--\ref{subsec:exp2}, all placement policies use the same all-\texttt{Linear} persistent layout and incur conversion or relayout costs. WLA is enabled only in Section~\ref{subsec:exp3}. Scheduler hyperparameters, modeling overheads, and overhead-sensitive settings are documented in the public \href{https://github.com/YIAI-02/TriForm/blob/micro26_pieak_final/docs/EXPERIMENT_HYPERPARAMETERS.md}{experiment-hyperparameter documentation} and \href{https://github.com/YIAI-02/TriForm/blob/micro26_pieak_final/README.md}{DOPS README}.
}

\subsection{Verification of Scheduling Benefits and Speedup Analysis}
\label{subsec:exp1}

Figure~\ref{fig:exp_verification_npu} presents representative verification results for two model families on \texttt{HP32}. For clarity, we show batch size 16 for Llama-7B and batch size 8 for Qwen-1.8B. Each model covers \textit{prefill} lengths of 128 and 1024 and \textit{decode} lengths of 128, 512, and 1024. For each workload, we report both absolute latency and end-to-end speedup over \texttt{PD}. 
{\color{black}Each stacked bar is annotated with a prefill-to-decode latency ratio normalized as $T_{\mathrm{prefill}}:T_{\mathrm{decode}}=1:r$.}
In this experiment, all weight blocks are stored in the \texttt{Linear} layout so that the comparison isolates the effect of scheduling. 

For Llama-7B, the simulated \framework speedup over \texttt{PD} ranges from 1.10$\times$ to 1.48$\times$; for Qwen-1.8B, it ranges from 1.89$\times$ to 2.43$\times$. Validation using NPU hardware snippets and the PIM simulator produces speedups of up to 1.47$\times$ and 2.41$\times$, respectively. Gains are observed in all 6 evaluated configurations for both model families, whereas several static baselines remain sensitive to workload shape and sometimes underperform \texttt{PD}.
\begin{figure}[t]
	\centering
	\includegraphics[width=\linewidth]{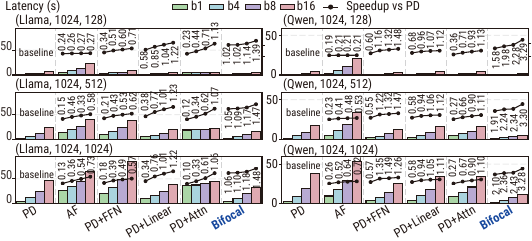}
	\caption{Speedup over \texttt{PD} across batch sizes 1, 4, 8, and 16 for Llama-7B and Qwen-1.8B.}
	\label{fig:exp_batch}
\end{figure}

\begin{figure}[t]
	\centering
	\includegraphics[width=\linewidth]{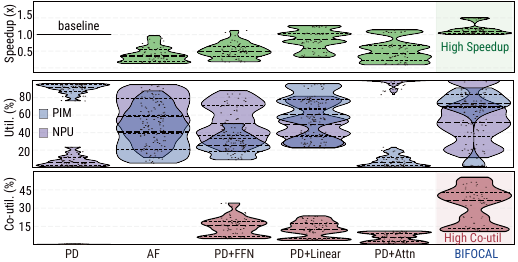}
		\caption{Distributions of speedup, average device utilization, and co-utilization on Llama-7B.}
	\label{fig:exp1_util}
\end{figure}
We further evaluate a $4\times4\times4$ sweep (64 workloads per model) over \textit{prefill} lengths \{128, 512, 1024, 2048\}, \textit{decode} lengths \{128, 256, 512, 1024\}, and batch sizes \{1, 4, 8, 16\}. \textsc{Bifocal} achieves geometric-mean speedups of 2.23$\times$, 1.36$\times$, and 1.44$\times$ on Qwen-1.8B, Qwen-7B, and Qwen-14B with \texttt{HP32}; 1.20$\times$ and 1.25$\times$ on Llama-7B and Llama-13B with \texttt{HP64}; and 1.72$\times$ and 1.20$\times$ on Mixtral-8$\times$7B and Llama-70B with \texttt{HP128}. Each platform accommodates the model weights and peak KV cache; detailed results are available at \url{https://github.com/YIAI-02/TriForm/tree/micro26_pieak_final}.

{\color{black}Figure~\ref{fig:exp1_histogram} analyzes the signed gap $\Delta$ between simulated and verified end-to-end speedups for the validated program snippets. We define $\Delta=\frac{S_{\mathrm{simulation}}-S_{\mathrm{verification}}}{S_{\mathrm{verification}}}\times 100\%$, where \(S_{\mathrm{simulation}}\) and \(S_{\mathrm{verification}}\) denote the simulated and verified speedups over \texttt{PD}, respectively.}
Across the two model families, the signed gap ranges from approximately $-4\%$ to $+6\%$.
{\color{black}
Because \textsc{Bifocal} relies on calibrated operator-level primitives, inaccurate performance models can degrade scheduling quality. For the validated snippets, the observed gap suggests sufficient fidelity for the scheduling decisions studied here.}

Figure~\ref{fig:exp_batch} compares speedup over \texttt{PD} across batch sizes 1, 4, 8, and 16 on Llama-7B and Qwen-1.8B at \textit{prefill}=1024 and three \textit{decode} lengths, including the low-batch regime emphasized by recent edge-deployment studies~\cite{seo2025facil,Li2025H2LLM,wang2023selfconsistency,li2025multisample,leviathan2023speculative}. At batch size 1, high modeled PIM bandwidth makes \texttt{PD} a strong baseline, yet \textsc{Bifocal} improves all six configurations. Its advantage grows with batch size as static stage separation becomes less effective.

Figure~\ref{fig:exp1_util} reports speedup over \texttt{PD}, average device utilization, and co-utilization. \textsc{Bifocal} has the highest median and upper-tail speedup in Figure~\ref{fig:exp1_util}(a). The instantaneous utilizations in Figure~\ref{fig:exp1_util}(b) are
$\mathrm{Util}_{\mathrm{PIM}}(t)=\frac{\mathrm{Busy}_{\mathrm{PIM}}(t)}{\mathrm{Num}_{\mathrm{PIM}}}$,
$\mathrm{Util}_{\mathrm{NPU}}(t)=\frac{\mathrm{Busy}_{\mathrm{NPU}}(t)}{\mathrm{Num}_{\mathrm{NPU}}}$.
Here, $\mathrm{Busy}_d(t)$ and $\mathrm{Num}_d$ are the busy and total device counts of type $d$.

\begin{figure}[t]
  \centering
  \includegraphics[width=\linewidth]{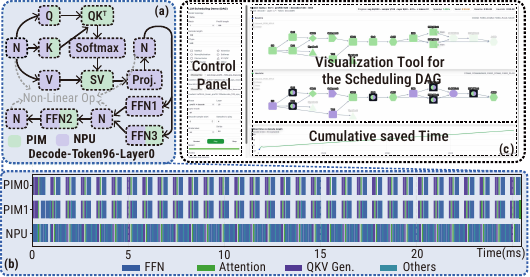}
    \caption{A case study of dynamic operator placement. (a) The operator-to-device mapping at token 96. (b) Timeline of device usage. (c) Our visualization tool for replaying the simulated schedule.}
  \label{fig:experiment_case}
\end{figure}

In Figure~\ref{fig:exp1_util}(c), we show the co-utilization defined as
\begin{equation}
\mathrm{CoUtil} = \frac{1}{T}\int_{0}^{T}\min\!\left(\mathrm{Util}_{\mathrm{PIM}}(t),\mathrm{Util}_{\mathrm{NPU}}(t)\right)\,dt
\end{equation}
where \(T\) denotes the makespan of the phase under consideration. This metric captures the time-averaged overlap between normalized PIM and NPU utilization. By definition, \(\mathrm{CoUtil}\) is upper-bounded by the smaller of the two standalone average utilizations; in practice, inter-operator dependencies together with synchronization and communication overheads in heterogeneous execution can make the gap substantial.

Across 64 Llama-7B cases, \framework does not always maximize NPU or PIM utilization in isolation, yet it consistently achieves the highest co-utilization. Dynamic assignment is effective because it selects the device that best matches the current global schedule, rather than simply the device with the shortest standalone execution time. Some static policies keep one device busy, yet still deliver limited speedup because the overlap between devices remains poor.

Figure~\ref{fig:experiment_case} presents an example for Llama-7B with \textit{prefill}=128, \textit{decode}=512, and batch size 16. Figure~\ref{fig:experiment_case}(a) visualizes the mapping at layer 0 of token 96, where \framework already moves the KV-cache-intensive portion of the attention path to PIM and offloads part of the FFN to increase overlap without creating an excessively long critical path. Figure~\ref{fig:experiment_case}(b) shows the PIM and NPU execution timeline for the decoding step that generates token 96 across the full set of layers, illustrating the high co-utilization between the two devices. Figure~\ref{fig:experiment_case}(c) shows a screenshot of our DAG-visualization tool. 
{\color{black}
Additional prefill and selected decode-token traces are provided in \url{https://github.com/YIAI-02/TriForm}; video of \url{https://youtu.be/Ya_oMCyYno0} shows a representative schedule replay with per-operator assignments and communication events.
}

\begin{figure}[t]
  \centering
  \begingroup
  \setlength{\fboxrule}{0pt}
  \setlength{\fboxsep}{0.5pt}

  \fcolorbox{black}{white}{
    \includegraphics[
      width=0.98\linewidth
    ]{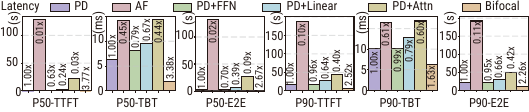}%
  }
  \endgroup

  \caption{\rev{Trace-driven BurstGPT serving on \texttt{HP32}.}}
  \label{fig:experiment_burstrequest}
\end{figure}

{\color{black}
To test realistic request dynamics, we replay 500 sampled BurstGPT requests with Qwen-1.8B on \texttt{HP32}, scaling arrival timestamps to create serving backpressure. The requests have mean input/output lengths of 708/286 tokens. In this experiment, $B_{\mathrm{token}}$ is disabled because realized output length is unknown at admission time. Requests are served in first-come-first-served (FCFS) order with a maximum serving batch size of 4 and a 10~ms batch timeout. Compared with the same static baselines, \textsc{Bifocal} improves median and tail TTFT, time-between-token (TBT), and end-to-end latency in Figure~\ref{fig:experiment_burstrequest}, confirming that \framework's dynamic scheduling benefits persist beyond fixed synthetic workloads.
}


\subsection{\color{black}Ablation Study of DOPS Scheduling}
\label{subsec:exp4}
\begin{figure}[t]
    \centering

    \begingroup
    \setlength{\fboxrule}{0pt}
    \setlength{\fboxsep}{0.5pt} 
    \fcolorbox{black}{white}{%
    \begin{minipage}{\dimexpr\columnwidth-2\fboxsep-2\fboxrule\relax}
        \centering

        \begin{minipage}[t]{0.56\linewidth}
            \vspace{0pt}
            \centering

            \resizebox{\linewidth}{!}{
            \begin{tabular}{lcccc}
                \toprule
                \multirow{2}{*}{\textbf{Variant}} &
                \multicolumn{2}{c}{\textbf{Near-focus}} &
                \multicolumn{2}{c}{\textbf{Far-focus}} \\
                \cmidrule(lr){2-3}
                \cmidrule(lr){4-5}
                &
                \textsc{EFT} &
                $\widehat{T}_{H}$ &
                $B_{\mathrm{phase}}$ &
                $B_{\mathrm{token}}$ \\
                \midrule
                \textsc{EFT-only}
                    & \cmark & -- & -- & -- \\
                \textsc{Bifocal} w/o LA
                    & \cmark & -- & \cmark & \cmark \\
                \textsc{Bifocal} w/o Phase
                    & \cmark & \cmark & -- & \cmark \\
                \textsc{Bifocal} w/o Token
                    & \cmark & \cmark & \cmark & -- \\
                \textsc{Bifocal}
                    & \cmark & \cmark & \cmark & \cmark \\
                \bottomrule
            \end{tabular}
            }

            \vspace{0.5em}
            \footnotesize
            \cmark: enabled;
			\footnotesize
            Note: LA = DAG-window lookahead term.
        \end{minipage}
        \hfill
        \begin{minipage}[t]{0.4\linewidth}
            \vspace{0pt}
            \centering
            \includegraphics[width=\linewidth]{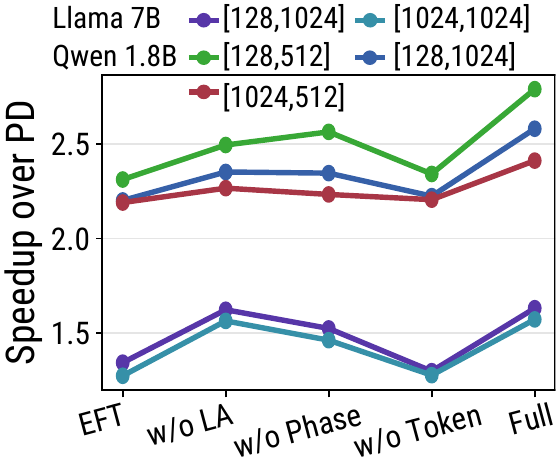}
        \end{minipage}

    \end{minipage}%
    }
    \endgroup

    \caption{\rev{Bifocal scheduler component ablation on selected representative workloads.}}
    \label{fig:ablation}
\end{figure}

{\color{black}
To isolate \textsc{Bifocal}'s gains, we ablate its score components on \texttt{HP32} using five Qwen-1.8B and Llama-7B representative workloads. Figure~\ref{fig:ablation} lists the enabled components in the left panel and reports speedup over \texttt{PD} in the right panel.

EFT-only captures much of the gain because it accounts for predecessor completion time, communication latency, device availability, device execution time, and cache-induced weight reload. The phase-reuse term consistently matters: removing it increases latency by 7.0\%--10.0\% by weakening weight-device affinity across the \textit{prefill}-\textit{decode} boundary and successive decode steps. The token-amortization term is critical for long decode; removing it causes 16.0\%--25.6\% loss because one-time migration or reload costs can be amortized over future tokens. The DAG-window lookahead term especially helps Qwen-1.8B: removing it adds 6.4\%--11.8\% latency, showing that successor awareness avoids downstream contention.}
\begin{table*}[tb]
    \begingroup
    \captionsetup{labelfont={color=black},textfont={color=black}}
    \color{black}
    \centering
    \scriptsize
    \setlength{\tabcolsep}{1.6pt}
    \renewcommand{\arraystretch}{1.22}
    \caption{Scheduler functionality comparison for PIM-accelerated LLM inference systems. Only scheduler-side functions are compared.}
    \label{tab:compared_works}
      \begin{tabularx}{\textwidth}{
        >{\raggedright\arraybackslash}m{1.8cm}|
        *{6}{>{\raggedright\arraybackslash}X}
        >{\raggedright\arraybackslash}m{2.5cm}
    }
        \noalign{\hrule height 1pt}
        \textbf{Method} &
        \textbf{AttAcc~\cite{park2024attacc}} &
        \textbf{IANUS~\cite{seo2024ianus}} &
        \textbf{FACIL~\cite{seo2025facil}} &
        \textbf{PAISE~\cite{lee2025paise}} &
        \textbf{PAPI~\cite{he2025papi}} &
        \textbf{PIMoE~\cite{wu2025pimoe}} &
        \textbf{DOPS} \\
        \hline
        \textbf{Scheduler type$^{(1)}$} &
        Static; offline &
        Static; offline &
        Static; offline &
        Dynamic; offline &
        Dynamic; online &
        Dynamic; online &
        \textbf{Dynamic; offline} \\
        \hline
        \textbf{Target stages and operators} &
        \textit{decode}:\newline Attn &
        \textit{prefill}+\textit{decode}:\newline Linear + Attn &
        \textit{decode}:\newline Linear &
        \textit{decode}:\newline Linear + Attn &
        \textit{decode}:\newline Linear &
        \textit{decode}:\newline MoE FFN &
        \textbf{\textit{prefill}+\textit{decode}}:\newline \textbf{all ops + comm.} \\
        \hline
        \textbf{Allocation basis$^{(2)}$} &
        Fixed-rule &
        Local-cost &
        Fixed-rule &
        Local-cost &
        Local-cost &
        Local-cost &
        \textbf{Global-timeline} \\
        \hline
        \textbf{Workload$^{(3)}$} &
        Dense &
        Dense &
        Dense &
        Dense &
        Dense &
        MoE &
        \textbf{ Dense + MoE} \\
        \hline
        \textbf{Extensibility$^{(4)}$} &
        Low&
        Medium &
        Low &
        Medium &
        Medium &
        Medium &
        \textbf{High} workload/hw. modular \\
        \noalign{\hrule height 1pt}
    \end{tabularx}
    \vspace{2pt}
    \begin{minipage}{\textwidth}
        \scriptsize
        \textit{Notes.}
        $^{(1)}$ Following the terminology introduced in Section~\ref{sec:intro}.
        $^{(2)}$ Fixed-rule = pre-defined operator-to-device mapping; Local-cost = consider only the metric of the current operator on a given device; Global-timeline = ready-DAG-task scoring with global horizon.
        $^{(3)}$ dense = no sparse expert routing.
        $^{(4)}$ Low = tied to a specific workload/hardware; Medium = reusable only within a restricted space; High = modular abstractions that can incorporate new operators or devices.
    \end{minipage}
    \endgroup
\end{table*}


\subsection{Hardware Scaling \& Marginal Returns}
\label{subsec:exp2}

\begin{figure}[t]
	\centering
	\includegraphics[width=\linewidth]{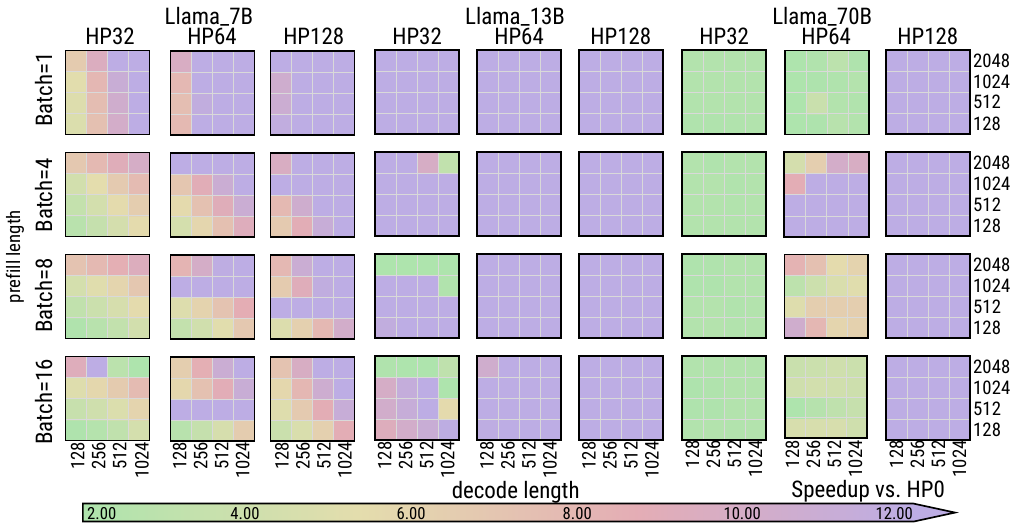}
	\caption{Speedup of \texttt{HP32}, \texttt{HP64}, and \texttt{HP128} compared to \texttt{HP0} across Llama-7B, Llama-13B, and Llama-70B.}
	\label{fig:exp2_heatmap}
\end{figure}

To evaluate how additional PIM capacity affects performance on the NPU--PIM platform for a given workload, we scale the hardware budget from \texttt{HP0} to \texttt{HP32}, \texttt{HP64}, and \texttt{HP128}, and report speedup over \texttt{HP0}. In this experiment, all configurations use the \textsc{Bifocal} scheduler, and weight blocks are stored in the linear layout. Figure~\ref{fig:exp2_heatmap} shows the resulting speedups for Llama-7B, Llama-13B, and Llama-70B across 64 combinations of batch size, \textit{prefill} length, and \textit{decode} length, where the color encodes the magnitude of the speedup relative to \texttt{HP0}. 
{\color{black}Additional PIM increases bandwidth and capacity but may also add communication/synchronization pressure, so observed benefits depend on \textsc{Bifocal}'s ability to distribute work across the NPU and PIM devices.}
Within each $4\times4$ block, corresponding to one batch size, the same PIM-capacity configuration generally yields larger speedup as the \textit{decode} length increases, moving right along the x-axis, and as the \textit{prefill} length increases, moving up along the y-axis. The main exceptions appear in several blocks with \textit{prefill} length 2048. In this regime, the KV cache occupies a substantial fraction of the PIM capacity, which weakens the benefit of accelerating weight-intensive operators.

\begin{figure}[t]
	\centering
	\includegraphics[width=\linewidth]{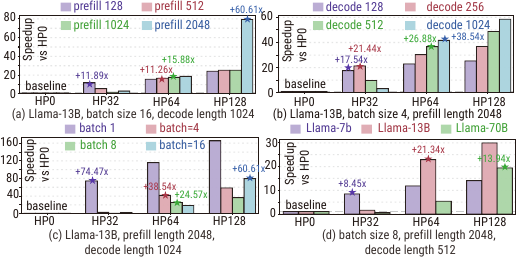}
	\caption{Marginal returns with hardware scaling.}
	\label{fig:exp2_marginal}
\end{figure}
Figure~\ref{fig:exp2_marginal} further explains the marginal return of each additional PIM budget. The bars represent speedup relative to \texttt{HP0}, while the starred point marks the configuration with the highest marginal payoff, namely, the capacity point whose incremental speedup over the previous smaller PIM configuration is the largest (not the highest absolute speedup). Although the overall trend is consistent, i.e., adding PIM is generally beneficial, the marginal gain is not universal. It is jointly determined by \textit{prefill} length, \textit{decode} length, batch size, and model size. Figure~\ref{fig:exp2_marginal}(a--d) presents 4 representative cases, where three factors are fixed while the remaining one varies to study its effect on the marginal return. 

\textbf{Insight 1}: \uline{Longer \textit{prefill} lengths shift the highest-payoff PIM point toward larger capacities.} As illustrated in Figure~\ref{fig:exp2_marginal}(a), under Llama-13B with batch size 16 and a fixed decode length of 1024, increasing the prefill length from 128 to 2048 shifts the strongest marginal gain from the \texttt{HP0}$\rightarrow$\texttt{HP32} increment to the \texttt{HP64}$\rightarrow$\texttt{HP128} increment. Correspondingly, the largest incremental gain increases from 11.89$\times$ to 60.61$\times$. This trend is expected because a longer prefill enlarges the context seen by subsequent decode steps, thereby increasing both KV-cache traffic and the memory cost of attention.

\textbf{Insight 2}: \uline{Longer \textit{decode} lengths induce the same rightward shift.} As illustrated in Figure~\ref{fig:exp2_marginal}(b), using Llama-13B with batch size 4 and a fixed prefill length of 2048, extending the decode length from 128 to 1024 shifts the largest marginal benefit from \texttt{HP0}$\rightarrow$\texttt{HP32} increment to \texttt{HP32}$\rightarrow$\texttt{HP64} increment, while the corresponding largest incremental gain increases from 17.54$\times$ to 38.54$\times$. Because KV-cache accesses and attention-related memory traffic are repeated over more generation steps, a larger PIM budget can amortize its cost more effectively.

\textbf{Insight 3}: \uline{A larger batch size can delay the payoff of increasing PIM capacity.} As illustrated in Figure~\ref{fig:exp2_marginal}(c), for Llama-13B with \textit{prefill} length fixed at 2048 and \textit{decode} length fixed at 1024, increasing the batch size from \textbf{1} to \textbf{4} shifts the largest marginal gain from \texttt{HP0}$\rightarrow$\texttt{HP32} increment to the \texttt{HP64}$\rightarrow$\texttt{HP128} increment. Intuitively, a larger batch improves NPU utilization in the \texttt{HP0} baseline, so the benefit of provisioning more PIM capacity emerges later.

\textbf{Insight 4}: \uline{Increasing model size pushes the optimal marginal-return point toward larger PIM capacities.} As illustrated in Figure~\ref{fig:exp2_marginal}(d), with batch size 8, prefill length 2048, and decode length 512 fixed, moving from Llama-7B to Llama-13B and then to Llama-70B shifts the strongest marginal gain from the \texttt{HP0}$\rightarrow$\texttt{HP32} increment to the \texttt{HP64}$\rightarrow$\texttt{HP128} increment. Larger models carry more weights, so an undersized PIM is more likely to trigger repeated LRU replacement and additional communication, making larger-capacity PIM configurations increasingly valuable.

\begin{figure}[tb]
  \centering
  \includegraphics[width=\linewidth]{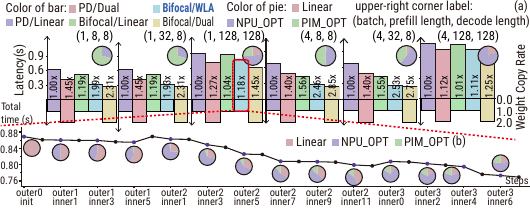}
  \caption{(a) Latency and speedup of \texttt{Bifocal/WLA} on Llama-7B. Bars report latency, annotations report speedup over \texttt{PD/Linear}, and pie charts show layout composition. The weight-copy rate is the normalized number of stored weight copies. (b) WLA optimization trace for a representative case.}
  \label{fig:exp3}
\end{figure}

\subsection{Effectiveness of WLA}
\label{subsec:exp3}

{\color{black}
In this experiment, we use \texttt{HP32} and evaluate the models that can be accommodated on this platform under the target workload settings: Llama-7B, Qwen-1.8B, Qwen-7B, and Qwen-14B. Their weights are partitioned into 56, 49, 49, and 70 blocks, respectively.} We sweep \textit{prefill} and \textit{decode} lengths over \{8, 32, 64, 128, 1024\}, and batch sizes over \{1, 4, 8\}.
Compared to \texttt{Bifocal/Linear}, \texttt{Bifocal/WLA} reduces latency by 8.3\%--40.4\%, with greater benefits at shorter sequence lengths where layout overhead dominates.

Across the full sweep available at \url{https://github.com/YIAI-02/TriForm}, \texttt{Bifocal/WLA} achieves geometric-mean speedups of 2.12$\times$ over \texttt{PD/Linear} and 1.32$\times$ over \texttt{Bifocal/Linear} on Qwen-1.8B, with consistent gains on larger models. These results show that WLA nearly matches the dual-copy baseline without doubling persistent weight storage.

The sweep also reveals format preferences: at low batch sizes and short decode lengths, \texttt{PIM\_OPT} blocks dominate to avoid high one-time conversion costs. As batch size and sequence length grow, \texttt{NPU\_OPT} becomes preferred to avoid repeated NPU-side transformations. Only near-balanced cases retain a few \texttt{Linear} blocks.
Figure~\ref{fig:exp3}(b) shows rapid convergence over iterations: from an all-\texttt{Linear} start, coarse blockwise moves are made toward the load-favored format, followed by limited refinement. Although latency is not strictly monotonic, a lower-latency mixed layout is typically reached within three outer iterations.

\begin{figure}[t]
	\centering
	\setlength{\fboxrule}{1pt}
	\setlength{\fboxsep}{0pt} 
    \includegraphics[width=0.98\linewidth]{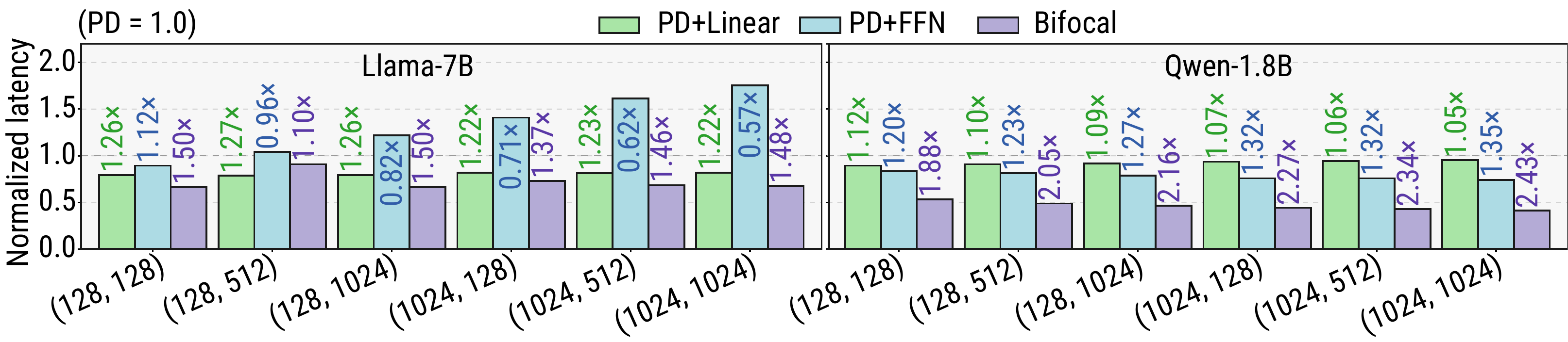}
    \caption{\rev{Prior-scheduler-inspired baselines on \texttt{HP32} for representative Llama-7B and Qwen-1.8B workloads.}}
	\label{fig:exp6_runtime_compare}
\end{figure}

{\color{black}

\section{Related Work}

\textbf{Scheduling on heterogeneous platforms.} Recent LLM-oriented PIM systems cover different operator--device placement points, as summarized in Table~\ref{tab:compared_works}. These policies are useful but restricted, while \framework searches a broader space to select a workload- and hardware-aware plan.

\framework is orthogonal and complementary to two lines of prior work. First, \framework is complementary to hardware-specific mechanisms or data-layout transformations, such as near-memory data condenser to bridge sparse data layouts between NPU and PIM in PIMoE~\cite{wu2025pimoe}. \framework can consume such optimized substrates through its hardware abstraction and performance model. Second, \framework is complementary to runtime schedulers. Online mechanisms can adapt to instantaneous serving states~\cite{he2025papi,wu2025pimoe}. However, it must remain lightweight. In contrast, \framework performs optimization at deployment time and can afford a broader search over the full inference DAG. 

As shown in Figure~\ref{fig:exp6_runtime_compare}, when the PAPI- and PAISE-inspired offloading rules are transplanted into our common evaluation framework, their runtime thresholds reduce to deterministic placement policies for each fixed workload and hardware configuration. We instantiate them as the closest static policy classes in Table~\ref{tab:compared_methods}. Nevertheless, their policies remain valuable when the actual serving state deviates from the profiled configuration.}

{\color{black}
\textbf{Hardware simulators.} Existing hardware modeling and simulation tools  typically target a specific abstraction level or accelerator substrate~\cite{zhang2024llmcompass,ham2024onnxim,xu2018pimsim,forlin2022sim2pim,luo2023ramulator2}. 
Detailed simulators improve the fidelity of performance model estimates for a given hardware design, whereas \framework optimizes how a full LLM inference DAG should be mapped to heterogeneous devices under such estimates. In practice, these simulators can serve as offline calibration backends for our performance model, replacing or refining roofline-based and profile-based operator costs. Conversely, \framework provides a system-level optimization layer above them and extends their utility from isolated hardware evaluation to end-to-end scheduling for \platform LLM inference.
}

\section{Conclusion}
We develop \framework and formulate LLM inference on heterogeneous \platform systems as a coupled scheduling-and-layout problem covering operator placement, device contention, and persistent weight format. Across the studied model--platform pairings, \textsc{Bifocal} achieves 1.20$\times$--2.23$\times$ geometric-mean speedup over \texttt{PD}, while WLA adds 1.28$\times$--1.33$\times$ over \texttt{Bifocal/Linear}. On these edge-oriented systems, the results show the value of moving beyond coarse \textit{prefill}--\textit{decode} disaggregation and static roofline rules and support future hardware--software co-design.


\bibliographystyle{unsrtnat}
\bibliography{ref}

\end{document}